\documentclass[apj,iop]{emulateapj}
\usepackage[colorlinks,citecolor=blue,linkcolor=blue,urlcolor=blue]{hyperref}
\usepackage{bm}

\newcommand {\be} {\begin{equation}}
\newcommand {\ee} {\end{equation}}
\newcommand {\bea} {\begin{eqnarray}}
\newcommand {\eea} {\end{eqnarray}}

\begin{document}

\title{Kinetic simulations of the lowest-order unstable mode of relativistic magnetostatic equilibria}
\shorttitle{Kinetic simulations of magnetostatic equilibria}

\author{
Krzysztof~Nalewajko\altaffilmark{1,2,3},
Jonathan~Zrake\altaffilmark{2},
Yajie~Yuan\altaffilmark{2},
William~E.~East\altaffilmark{2},
Roger~D.~Blandford\altaffilmark{2}
}
\shortauthors{Nalewajko et~al.}

\altaffiltext{1}{Nicolaus Copernicus Astronomical Center, Bartycka 18, 00-716 Warsaw, Poland; {\tt knalew@camk.edu.pl}}
\altaffiltext{2}{Kavli Institute for Particle Astrophysics and Cosmology, SLAC National Accelerator Laboratory, Stanford University, 2575 Sand Hill Road M/S 29, Menlo Park, CA 94025, USA}
\altaffiltext{3}{NASA Einstein Postdoctoral Fellow}

\begin{abstract}
We present the results of particle-in-cell numerical pair plasma simulations of relativistic 2D magnetostatic equilibria known as the ``ABC'' fields.
In particular, we focus on the lowest-order unstable configuration consisting of two minima and two maxima of the magnetic vector potential.
Breaking of the initial symmetry leads to exponential growth of the electric energy and to the formation of two current layers, which is consistent with the picture of ``X-point collapse'' first described by Syrovatskii.
Magnetic reconnection within the layers heats a fraction of particles to very high energies.
After the saturation of the linear instability, the current layers are disrupted and the system evolves chaotically, diffusing the particle energies in a stochastic second-order Fermi process leading to the formation of power-law energy distributions.
The power-law slopes harden with the increasing mean magnetization, but they are significantly softer than those produced in simulations initiated from Harris-type layers.
The maximum particle energy is proportional to the mean magnetization, which is attributed partly to the increase of the effective electric field and partly to the increase of the acceleration time scale.
We describe in detail the evolving structure of the dynamical current layers, and report on the conservation of magnetic helicity.
These results can be applied to highly magnetized astrophysical environments, where ideal plasma instabilities trigger rapid magnetic dissipation with efficient particle acceleration and flares of high-energy radiation.
\end{abstract}


\section{Introduction}

Observations of high-energy photons from certain astrophysical sources --- such as blazars, gamma-ray bursts, pulsars, etc. --- often reveal dramatic energy dissipation and efficient particle acceleration on very short time scales.
Examples include rapid gamma-ray variability of blazars \citep[e.g.,][]{Aha07,Ale11,Sai13,Hay15}, radio galaxies \citep[e.g.,][]{Ale14}, gamma-ray bursts \citep[e.g.,][]{Abd09}, and the Crab pulsar wind nebula \citep{Tav11,Abd11,Bue12}.
In many cases, the environment of these events is thought to be a highly magnetized collisionless plasma --- AGN jets, pulsar wind nebulae, etc.
This suggests scenarios involving efficient localized dissipation of magnetic energy allowing for rapid particle acceleration \citep[e.g.,][]{Beg08,Gia09,Nal11,Uzd11,Cla12}.

It is important to consider the astrophysically realistic situations leading to efficient magnetic dissipation.
The foremost requirement is the localized reversal of the magnetic field lines.
Using the Harris-type current layer as initial condition presumes a highly synchronized global reversal of the magnetic field polarity with a steady supply of magnetized plasma.
Such initial condition includes a rather arbitrary structure on kinetic scales, the impact of which can probably be neglected only in sufficiently large simulations \citep{Sir14,Guo14,Wer16}.
In the context of pulsar wind nebulae, global reversals can be readily realized in the equatorial striped wind \citep{Cor90,Lyu01,Kir03}, and their consequences for particle acceleration and high-energy emission are actively investigated \citep{Sir11,Bat13,Zra15}.
In the context of astrophysical jets, it has long been hypothesized that such global reversals can take place \citep[e.g.,][]{Lov97}, however, a convincing demonstration of such scenario is yet to be made.
The alternative is that magnetic dissipation is triggered by jet instabilities, in particular by the current-driven kink modes \citep{Beg98}.
In such a case, interactions of large-scale magnetic structures (eddies) are expected to create many transient localized current sheets that facilitate magnetic dissipation \citep{ONe12,Miz12}.

Currently, most investigations focus on relativistic magnetic reconnection, and indeed, great progress has been made over the past few years, in large part due to increasingly powerful direct kinetic plasma simulations.
When starting from a uniform Harris-type current layer, it has been convincingly demonstrated that relativistic reconnection leads to very efficient particle acceleration.
The resulting energy distributions can be characterized as power-laws $N(\gamma) \propto \gamma^{-p}$ with broad cut-offs, and the slope $p$ depending mainly on the background plasma magnetization $\sigma$.
In the limit of $\sigma \gg 1$, the energy distributions become extremely hard, with $p \simeq 1-1.2$ \citep{Sir14,Guo14,Wer16}.
Such hard energy distributions mean that the maximum particle energy is limited by the initial average magnetic energy per particle.
Indeed, for sufficiently large systems, the characteristic cut-off energy scales like $\gamma_{\rm c} \propto \sigma$ \citep{Wer16}.

However, the dissipation efficiency of relativistic reconnection is limited by reconnection rates being of order $\mathcal{O}(0.1)$ \citep[e.g.,][]{Liu15}.
Our group proposed an idea of \emph{magnetoluminescence}, a generic dissipation process that allows for rapid and efficient conversion of electromagnetic energy into radiation \citep{Bla14,Bla15}.
With this end, we started to explore a range of novel magnetostatic equilibria.

Recently, some of us investigated a class of unstable magnetostatic equilibria known as the ``Arnold-Beltrami-Childress'' (ABC) fields, using relativistic magnetohydrodynamics (RMHD) and force-free electrodynamics (FF) codes \citep{Eas15}.
Here, we present the first results of the subsequent investigation of these equilibria with the kinetic particle-in-cell (PIC) code {\tt Zeltron}.
The simulations presented here are two-dimensional, with pair plasma, and they focus on the lowest-order unstable mode.
Nevertheless, we obtained a very rich physical picture including the evolution of the sheared current layers, regular and stochastic modes of particle acceleration, and rapid dissipation of magnetic energy.

In Section \ref{sec_setup}, we define the initial configuration and its implementation in the {\tt Zeltron} code.
In Section \ref{sec_res}, we present the results including the evolution of total energy components, conservation of magnetic helicity, the evolving structure of the current layers, evolution of the particle energy distribution, and analysis of individual energetic particles.
Our results are discussed in Section \ref{sec_dis}, and our conclusions are presented in Section \ref{sec_con}.

\section{Simulation setup}
\label{sec_setup}

We performed two-dimensional PIC numerical simulations using the {\tt Zeltron}
code\footnote{\url{http://benoit.cerutti.free.fr/Zeltron/}} \citep{Cer13}. The
core algorithms include a Finite-Difference-Time-Domain (FDTD) method for
advancing the electromagnetic fields on staggered grids \citep{Yee66}, the ``Boris
push'' for advancing the particles, a smoothing filtering of the electric fields,
and a Poisson solver for matching the electric fields with the charge density. The
radiation reaction implemented in this code was not used in these simulations.

Our simulations proceed from smooth magnetostatic equilibria.
They are kinetic generalizations of the MHD states whose gas pressure is uniform and Lorentz force $\bm{j} \times \bm{B}$ vanishes.
We focus on force-free equilibria satisfying the Beltrami condition $\bm\nabla \times \bm{B} = \alpha\bm{B}$, where $\alpha$ is a constant.
In the present study we use the following magnetic field
\bea
\label{eqn_ic}
B_x &=& B_0\left[\sin(\alpha_0(x+y))+\sin(\alpha_0(x-y))\right]/\sqrt{2}\,,
\nonumber
\\
B_y &=& B_0\left[\sin(\alpha_0(x-y))-\sin(\alpha_0(x+y))\right]/\sqrt{2}\,,
\\
B_z &=& B_0\left[\cos(\alpha_0(x+y))-\cos(\alpha_0(x-y))\right]\,,
\nonumber
\eea
where $\alpha_0 = 2\pi/L$ for linear domain size $L$.
This state has a wavelength that is smaller by a factor $\sqrt{2}$ than the domain scale,
and resembles the conventional ABC magnetic fields we used in previous studies, but has been rotated by $45^\circ$.
Translational symmetry of this state along $z$ is enforced throughout the evolution by the ``2.5D'' simulation scheme.
This state contains two minima and two maxima of the out-of-plane vector potential $A_z$.
Those extrema are centered on helical flux tubes aligned with the $z$-axis, which are linearly unstable to pairwise coalescence instability as we reported in \cite{Eas15}.
As such coalescence moves plasma toward a lower energy (longer wavelength) magnetic configuration, the initial state must possess a non-zero ``magnetostatic free energy''.
That is, by our definition, the magnetic
energy that could be removed while the frozen-in MHD condition ---
$\bm{E} = \bm{B}\times\bm\beta$, hence $\bm{E}\cdot\bm{B} = 0$
--- is maintained over all but infinitesimal volumes. Such evolution leads to
minimization of the magnetic energy as constrained by the global magnetic
helicity invariant in the sense of \cite{Tay74}. For general states having
a wavelength $\lambda_0$, the theoretical free energy fraction is in general
$f_{\rm B,0} = 1 - \lambda_0 / L$, which has the value $1 - 1/\sqrt{2} \approx
0.29$ for our setup. In other words, $29\%$ of the magnetic energy of our
initial condition could be used to energize particles if magnetic helicity is
conserved.

The current density $\bm{j} = -(c/\sqrt{2}L)\bm{B}$ of the setup in Equation
(\ref{eqn_ic}) is realized by endowing particles with a locally anisotropic
momentum distribution, which we factorize into independent energy and angular
parts as
\be
f(\bm{u})\,{\rm d}^3u = f_1(\gamma,\Theta)\left(\frac{1+a_1\mu}{2}\right)\,{\rm d}\gamma\,{\rm d}\mu\,,
\ee
where $\bm{u} = \gamma\bm\beta$ is the dimensionless particle momentum, $\gamma = (1+u^2)^{1/2}$ is the dimensionless particle energy (Lorentz factor), $\bm\beta = \bm{v}/c$ is the dimensionless particle velocity. Function $f_1(\gamma,\Theta)$ is the Maxwell-J\"{u}ttner energy distribution for dimensionless temperature $\Theta = kT/mc^2$; $\mu = \cos\theta$, where $\theta$ is the polar angle measured from the unit vector aligned with the required local current density vector. Parameter $a_1 \le 1$ is the dipole moment of the local angular distribution of paricle momenta, which determines the local average drifting speed $\left<\beta_{\rm d}\right> = (a_1/3)\left<\beta\right>$, where $\left<\beta\right>$ is the average speed that is a function of $\Theta$.

\begin{table*}
\centering
\caption{Parameters of the simulation runs. Two characteristic magnetization values are reported: $\sigma_{\rm cold}$ and $\sigma_{\rm hot}$. Parameter $\tilde{a}_1$ is the characteristic constant of the dipole moment of particle momentum distribution. Parameter $\tau$ is the e-folding growth rate of the total electric energy.
Parameter $\epsilon_{\rm diss} = f_{\rm B}/f_{\rm B,0}$ describes the global efficiency of magnetic energy conversion, where $f_{\rm B} = 1-E_{\rm B,fin}/E_{\rm B,ini}$ is the measured conversion efficiency, and $f_{\rm B,0} = 1-2^{-1/2} = 29.3\%$ is the theoretical expectation for relaxation to the isotropic Taylor state.
Parameters $f_{\rm n}$ and $f_{\rm e}$ describe the number and energy fractions contained in the high-energy particle distribution tail at $ct/L \simeq 10$. Parameter $p$ is the index of the power-law energy distribution tail estimated at $ct/L \simeq 10$. Parameter $\gamma_{\rm max}$ is the maximum particle energy measured at the $10^{-3}$ level of the normalized energy distribution at $ct/L \simeq 10$.
}
\vskip 1ex
\label{table_runs}
\begin{tabular}{lrrrrrrrrrr}
\hline\hline
Run & $L/\rho_0$ & $\tilde{a}_1$ & $\sigma_{\rm cold}$ & $\sigma_{\rm hot}$ & $c\tau/L$ & $\epsilon_{\rm diss}$ & $f_{\rm n}$ & $f_{\rm e}$ & $p$ & $\gamma_{\rm max}$ \\
\hline
 s07L400 &  400 & 1/4 &  3.4 &  0.7 & 0.33 & 0.77 & 0.03 & 0.11 & 3.3 & 105 \\ 
 s07L800 &  800 & 1/8 &  3.4 &  0.7 & 0.36 & 0.78 & 0.04 & 0.12 & 3.7 &  95 \\ 
\hline
 s14L400 &  400 & 1/2 &  6.8 &  1.4 & 0.23 & 0.86 & 0.07 & 0.23 & 3.1 & 175 \\ 
 s14L800 &  800 & 1/4 &  6.8 &  1.4 & 0.25 & 0.80 & 0.06 & 0.21 & 2.9 & 180 \\ 
s14L1600 & 1600 & 1/8 &  6.8 &  1.4 & 0.27 & 0.78 & 0.06 & 0.20 & 3.1 & 255 \\ 
\hline
 s27L800 &  800 & 1/2 & 13.6 &  2.7 & 0.20 & 0.84 & 0.09 & 0.30 & 2.9 & 380 \\ 
s27L1600 & 1600 & 1/4 & 13.6 &  2.7 & 0.21 & 0.81 & 0.10 & 0.38 & 2.7 & 480 \\ 
\hline
s55L1600 & 1600 & 1/2 & 27.2 &  5.5 & 0.17 & 0.83 & 0.17 & 0.59 & 2.5 & 790 \\ 
s55L3200 & 3200 & 1/4 & 27.2 &  5.5 & 0.18 \\ 
\hline\hline
\end{tabular}
\end{table*}

For pair plasma, the total current density module is $j = ecn\left<\beta_{\rm d}\right>$, where $n$ is the number density of both electrons and positrons.
Combining this with previous relations, we obtain an expression for equilibrium value of the pair number density:
\be
n = \frac{3\sqrt{2}B_0}{2e\tilde{a}_1\left<\beta\right>L}\,,
\ee
where $\tilde{a}_1 = (B_0/B)a_1 \le 1/2$ is a constant equal to a characteristic value of the dipole moment (note that $a_1(x,y) \propto B(x,y)$).
The upper limit imposed on the dipole moment introduces a lower limit on the particle number density, and hence an upper limit on the magnetization of the simulated system. The characteristic value of the ``cold magnetization'' can be expressed as:
\be
\sigma_{\rm cold} = \frac{B_0^2}{4\pi nm_{\rm e}c^2} = \frac{\tilde{a}_1\left<\beta\right>}{6\sqrt{2}\pi}\left(\frac{L}{\rho_0}\right)\,,
\ee
where $\rho_0 = m_{\rm e}c^2/(eB_0)$ is the nominal gyroradius. Hence, for the maximum value of the dipole moment, the magnetization scales linearly with the system size $(L/\rho_0)$. For isotropic distributions, the cold magnetization is related to the plasma skin depth $d_{\rm e} = [m_{\rm e}c^2/(4\pi e^2n)]^{1/2} = \sigma_{\rm cold}^{1/2}\rho_0$.
We also define the ``hot magnetization'' as $\sigma_{\rm hot} = B_0^2/(4\pi w)$, where $w = \left<\gamma(1+\beta^2/3)\right>nm_{\rm e}c^2$ is the specific enthalpy. However, one should note that even the initial magnetization is not uniform due to the non-uniformity of $|\bm{B}|$.

We have performed several simulations for different values of $(L/\rho_0)$ and $\tilde{a}_1$, as reported in Table \ref{table_runs}. Common parameter values are $B_0 = 1\;{\rm G}$ and $\Theta = 1$. The numerical resolution is $\Delta x = \Delta y = \rho_0/2.56$, and the total number of particles per cell is 128.

\section{Results}
\label{sec_res}

\subsection{Snapshot maps}

\begin{figure*}
\centering
\includegraphics[width=\textwidth]{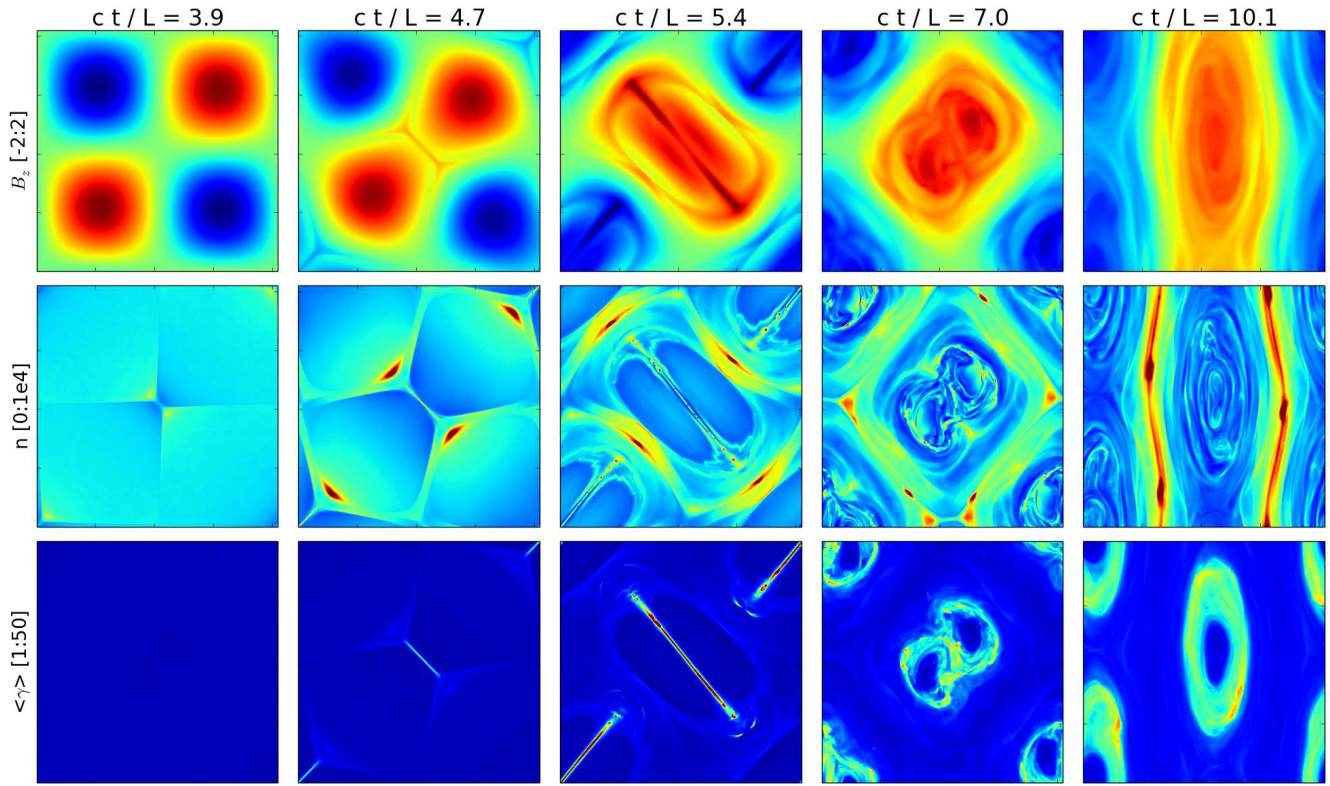}
\caption{Snapshots from the Run s55L1600. Time is normalized to the light-crossing time scale $L/c$. \emph{Top panels:} magnetic field component $B_z$; \emph{middle panels:} number density $n$ of electrons and positrons; \emph{bottom panels:} average Lorentz factor $\left<\gamma\right>$ of electrons and positrons.}
\label{fig_snapshots}
\end{figure*}

Figure \ref{fig_snapshots} shows the distribution of magnetic field component $B_z$, particle number density $n$, and average particle Lorentz factor $\left<\gamma\right>$ at several time steps for Run s55L1600.
The initial condition includes two maxima (red) and two minima (blue) of $B_z$, which correspond to the extrema of the magnetic vector potential $A_z \propto B_z$.
By $ct/L = 3.9$, the initial symmetry of the setup is broken, as can be seen on the particle number density map.
Two current layers form, one in the center of the simulation domain, and one in the corners.
By $ct/L = 5.4$, the current layers grow in length, they accelerate particles to high energies, and they undergo a minor tearing instability.
This process can be characterized as the ``X-point collapse'' first investigated by \cite{Syr66}.
The sharp current layers seen at $ct/L = 5.4$ are disrupted, and the subsequent evolution of the magnetic domains proceeds through a series of oscillations forming complex, chaotic substructures.
However, in the final stage, we observe a clear separation between a dense cold plasma located along the magnetic domain boundaries, and a dilute hot plasma partially filling the domain interiors.

\subsection{Evolution of total energy}

\begin{figure*}
\centering
\includegraphics[width=\textwidth]{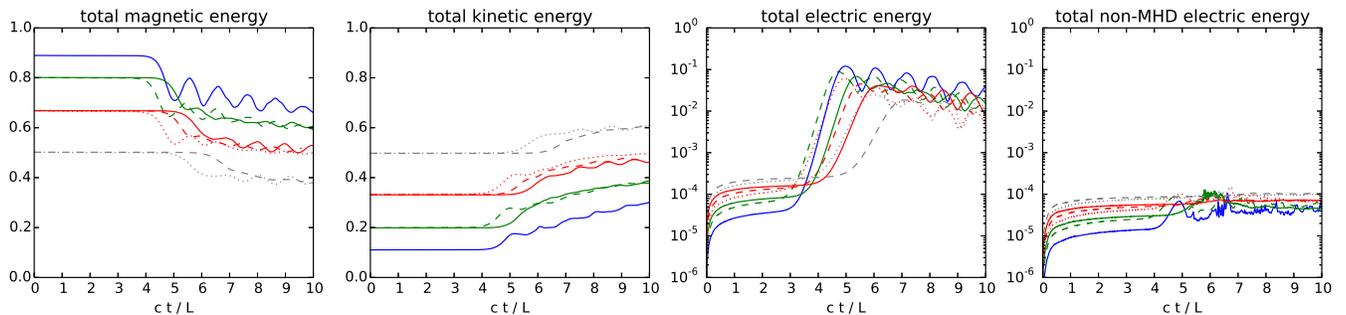}
\caption{Evolution of the total energy components compared for different runs. All values are normalized to the initial total energy. Line colors indicate the mean value of the hot magnetization: $\sigma_{\rm hot} = 0.7$ (gray), $\sigma_{\rm hot} = 1.4$ (red), $\sigma_{\rm hot} = 2.7$ (green), and $\sigma_{\rm hot} = 5.5$ (blue). Line types indicate the simulation domain size: $L = 400\rho_0$ (dotted), $L = 800\rho_0$ (dashed), and $L = 1600\rho_0$ (solid).}
\label{fig_total_energy}
\end{figure*}

Figure \ref{fig_total_energy} shows the time evolution of the main components of the total energy: magnetic, kinetic (including thermal), and electric.
The total energy in our simulations is conserved at the level better than $10^{-3}$, improving with increasing magnetization.
All simulations (except for s55L3200) are run for at least 10 light-crossing time scales, at which point the evolution of the total energy components is largely complete.
Parameters of the initial configuration determine the characteristic magnetization value, and hence the initial energy shared between the magnetic field and particles.

The global efficiency of magnetic dissipation is calculated as $\epsilon_{\rm diss} = f_{\rm B} / f_{\rm B,0}$, where $f_{\rm B} = 1 - E_{\rm B,fin}/E_{\rm B,ini}$, $E_{\rm B,ini}$ is the total magnetic energy at $t = 0$, and $E_{\rm B,fin}$ is the total magnetic energy at $ct/L \simeq 10$.
As we report in Table \ref{table_runs}, $\epsilon_{\rm diss} \approx 80\%$ for all simulation runs.
This indicates that magnetic relaxation toward the Taylor minimum is effectively complete by that time.
We note here that in 2.5D, the energy of the most relaxed state generally exceeds the Taylor minimum energy due to constraints on the change of magnetic topology that were not considered by Taylor, who intended to characterize fully 3D evolution.
Nevertheless, \cite{ZraEas16} found the additional constraints became significant only when $\lambda_0 \ll L$.
Since we have $\lambda_0 = L/\sqrt{2}$, nearly complete relaxation to the Taylor minimum energy state is to be expected in the cases presented here.
We also estimate the peak $e$-folding time scale of the total magnetic energy as $\sim 2.5L/c$.

The initial electric energy, at the level of $\sim 10^{-4}$ of the total energy, is determined by the residual charge density due to the finite number of particles per cell. At $ct/L \simeq 3-4$, the electric energy begins to grow exponentially. By $ct/L \simeq 5-6$, the growth of the electric energy saturates at the level of $\sim 0.1$ of the total energy. At about the same time, the magnetic energy decreases sharply, and the kinetic energy begins to increase. There is no simple energy transformation between magnetic and kinetic forms. Instead, the electric energy begins to oscillate, and this oscillation is damping over many light-crossing time scales. These oscillations are reflected more clearly in the evolution of the magnetic energy component, rather than of the kinetic component.

We have also calculated the evolution of non-ideal electric field parallel to the magnetic field $E_{\parallel B} = \bm{E}\cdot\bm{B}/|\bm{B}|$. Figure \ref{fig_total_energy} shows that the growth of the non-ideal electric energy is negligible when compared with the growth of the total electric energy. For low magnetizations it is difficult to distinguish the total non-ideal component from the noise.
In fact, non-ideal regions form distinct compact spatial structures, while most of the domain volume is consistent with ideal MHD at all times.

\begin{figure}
\centering
\includegraphics[width=\columnwidth]{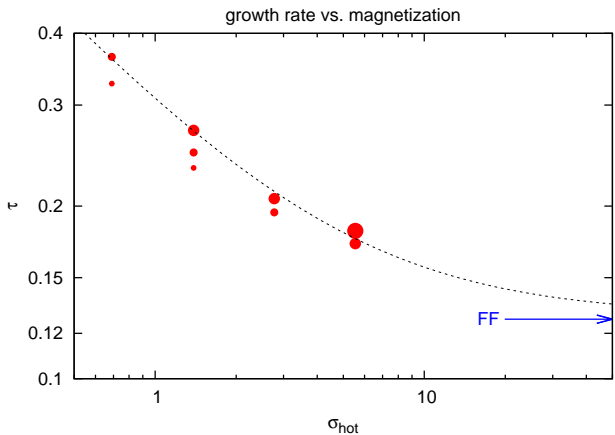}
\caption{Linear instability growth rate, defined as e-folding time scale of the total electric energy normalized to $L/c$, as function of the mean value of the hot magnetization. The point size indicates the simulation domain size $L$. The blue arrow indicates the asymptotic value measured in FF simulations \citep{Eas15}.}
\label{fig_growth_rates}
\end{figure}

We have measured the growth rates of the total electric energy during the linear instability phase as $e$-folding time $\tau$.
Figure \ref{fig_growth_rates} shows the dependence of $c\tau/L$ on the mean magnetization $\sigma_{\rm hot}$. For each value of $\sigma_{\rm hot}$, the growth rate increases systematically with the domain size, which suggests that we do not achieve a complete convergence. Considering only the growth rates measured for the largest simulation for each magnetization value, we fitted them with the following function $c\tau(\sigma_{\rm hot})/L = a/v_{\rm A}(s\sigma_{\rm hot})$, where $v_{\rm A}(\sigma) = [\sigma/(1+\sigma)]^{1/2}$ is the Alfv\'en speed. For the scaling parameters, we found $a \simeq 0.13$ and $s \simeq 0.21$.

\subsection{Magnetic helicity conservation}

\cite{ZraEas16} studied the decay of magnetic turbulence in the force-free limit in both 2D and 3D. They investigated the conservation of magnetic helicity $\mathcal{H} = \int \bm{A}\cdot\bm{B} {\rm d}V$, which is formally broken only by the presence of regions with non-ideal electric field, as $\dot \mathcal{H} = -2 \int \bm{E} \cdot \bm{B} {\rm d}V$ \citep[e.g.][]{Brandenburg2015}. In both 2D and in 3D, they reported that total magnetic helicity $\mathcal{H}$ was indeed conserved to a high precision, which improved with resolution. They also found that in 2D, $\mathcal{H}$ could be decomposed into a continuous distribution $d \mathcal{H}/d A_z$, each value of which is a separate invariant. Here we verify these findings in 2D PIC simulations for the first time.

\begin{figure*}[hb]
\centering
\includegraphics[width=0.33\textwidth]{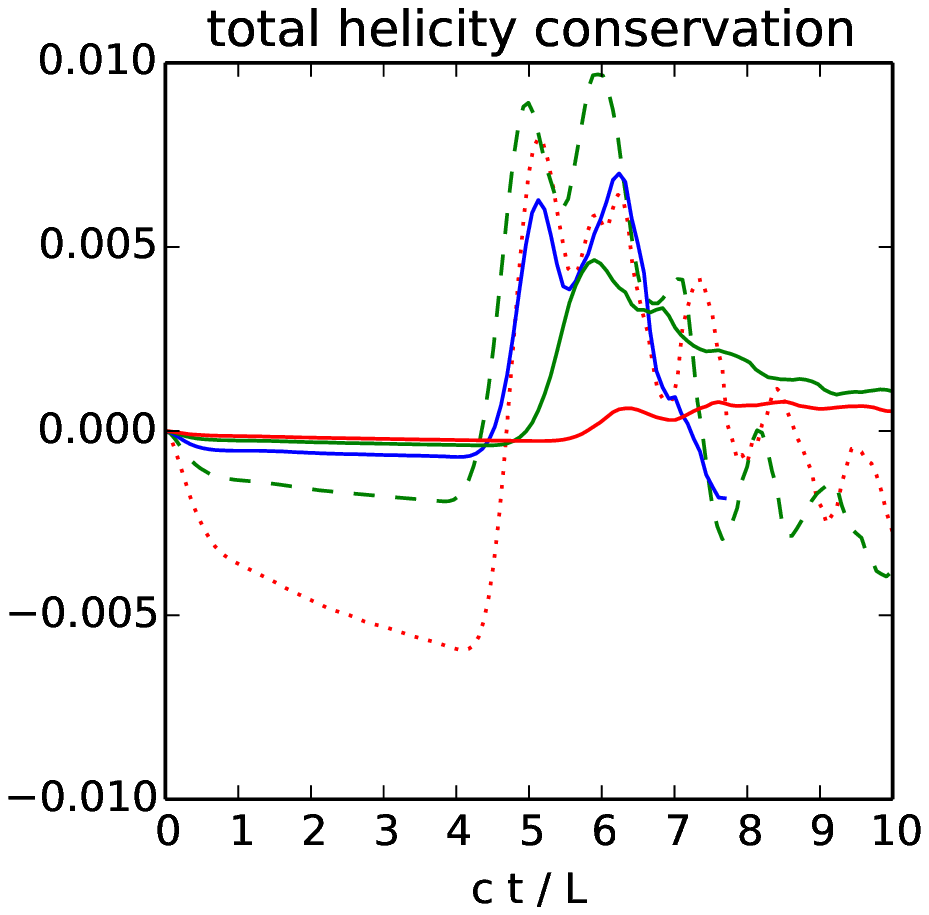}
\includegraphics[width=0.66\textwidth]{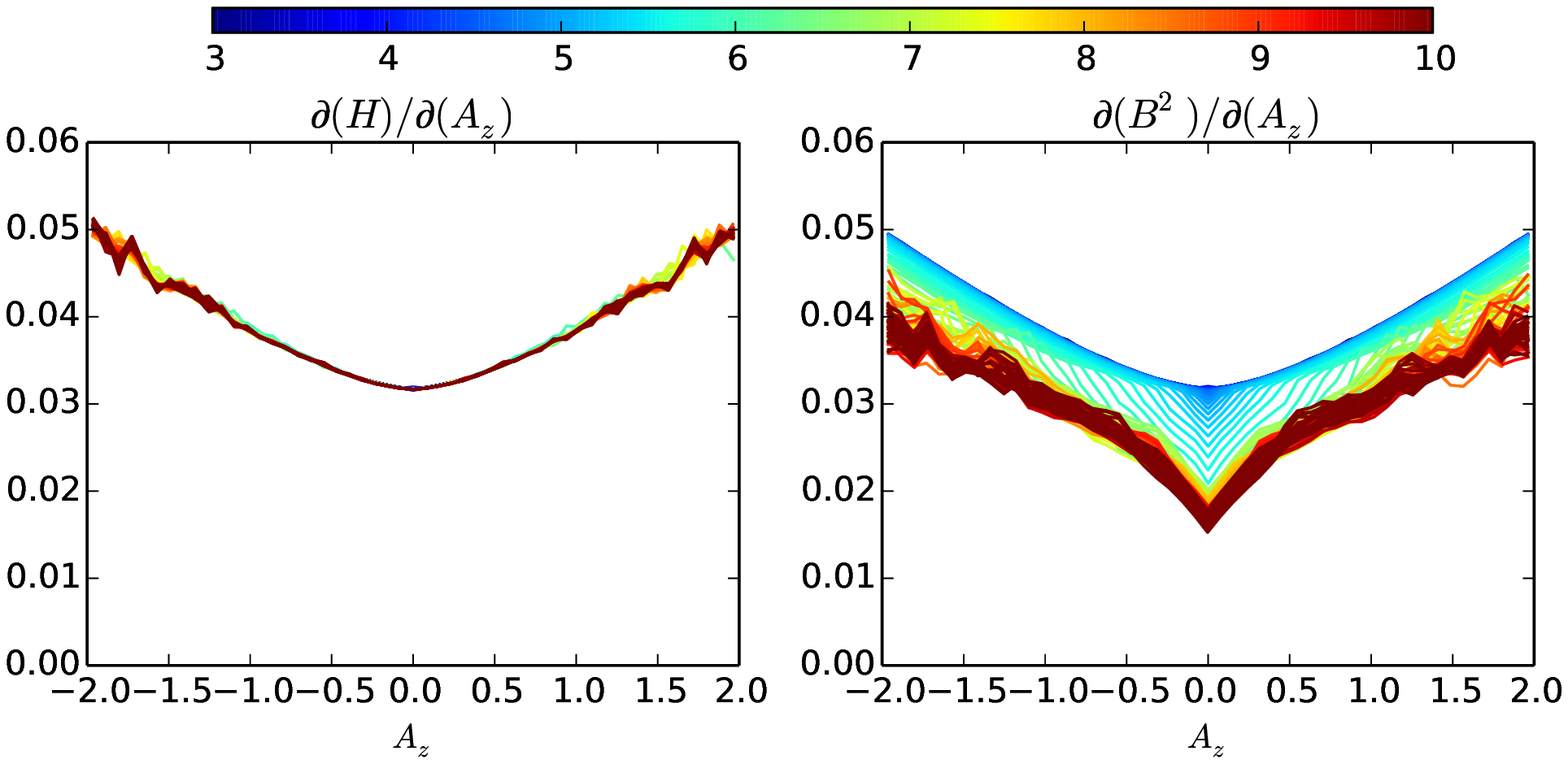}
\caption{
\emph{First panel (from the left):} the accuracy of total helicity conservation $\mathcal{H}(t)/\mathcal{H}(0)-1$ compared for several simulations. Line colors and types are the same as in Figure \ref{fig_total_energy}.
\emph{Second panel:} time evolution of the helicity profile $\partial\mathcal{H}/\partial A_z$ for Run s14L1600.
\emph{Third panel:} time evolution of the magnetic energy profile $\partial(B^2)/\partial A_z$ for Run s14L1600.
For the two right panels, the color scale indicates simulation time normalized to the light-crossing time scale $L/c$.
}
\label{fig_heli}
\end{figure*}

In Figure \ref{fig_heli}, we show that total magnetic helicity is conserved at the level better than $1\%$, improving with the decreasing particle momentum dipole moment $\tilde{a}_1$. As we will demonstrate later on, smaller dipole moment allows for higher particle density and better screening of the electric fields. Consequently, the volume fraction of the non-ideal field regions decreases, and this may explain better conservation of the magnetic helicity, in line with the results of \cite{ZraEas16} in the force-free limit. In Figure \ref{fig_heli}, we also show that while $d \mathcal{H}/d A_z$ is well conserved for all values of $A_z$, magnetic energy is dissipated during the linear instability phase for all values of $A_z$.

\subsection{Structure of the current layers}

Two current layers are formed during the late stages of the instability, at $4 \lesssim ct/L \lesssim 5.5$. They appear as thin structures characterized by high density and high average particle energy, they grow in length and undergo tearing instability until they are disrupted. These are also regions where non-ideal parallel electric fields are present, with $\bm{E}\cdot\bm{B} \ne 0$, and hence they enable efficient particle acceleration.

\begin{figure*}
\centering
\includegraphics[width=\textwidth]{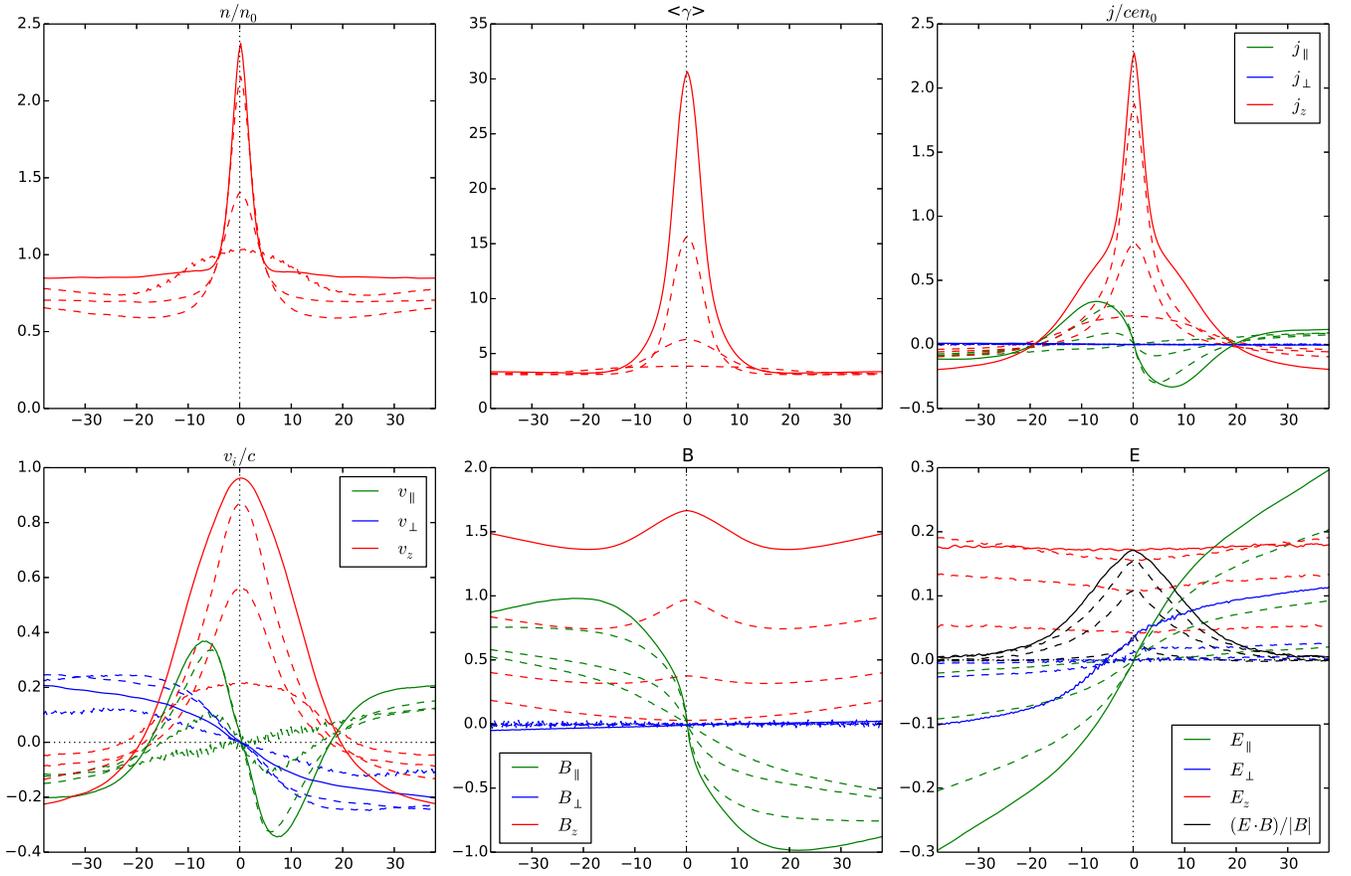}
\caption{Perpendicular profiles measured across the centroid of the current layer for Run s27L800 at several times: $ct/L = 3.8, 4.1, 4.4$ (\emph{dashed lines}), and $ct/L = 4.7$ (\emph{solid lines}). The panels show: (1) number density of electrons and positrons normalized to the initial value, (2) average energy of electrons and positrons, (3) three components of the current density normalized to the initial number density, (4) three components of the mean velocity of positrons, (5) three components of the magnetic field, (6) three components of the electric field and the $(\bm{E}\cdot\bm{B})/|\bm{B}|$ value. Subscript `$\parallel/\perp$' means the component parallel/perpendicular to the major axis of the current layer.}
\label{fig_prof_perp}
\end{figure*}

Since the current layers are not present in the initial configuration, it is interesting to characterize their perpendicular structure and evolution.
To this end, we first determined the exact location and orientation of the layers by fitting a two-dimensional Gaussian model to the spatial distribution of the average particle energy.
The time range when this can be done robustly is limited --- for $ct/L < 4$ the current layers are not detectable, and for $ct/L \gtrsim 5.5$ the current layers bend and eventually disappear.
We introduce a local coordinate system with parallel/perpendicular vectors measured along/across the major axis of the current layer.
As the location and orientation of the current layers changes slowly, this is repeated at every timestep of interest.
Next, we calculated perpendicular profiles of various parameters across the current layer centroid.
These profiles are shown in Figure \ref{fig_prof_perp} at 4 different times to illustrate a fairly complex evolution of the current layer.

The density profile develops a narrow spike, and the average particle energy profile shows a similar but somewhat broader structure.
On the other hand, the current density component $j_z$ shows a more complex profile with a narrow core and broader wings.
This appears to be a combination of the narrow density profile with the broad profile of the $z$-component of the drifting velocity.
We also observe a small component of alternating parallel current $j_{\parallel l}$ on the scale comparable with the velocity structure.
This current is of opposite sign to the background in-plane current that was already present in the initial configuration.
Evolution of magnetic field involves a gradual steepening of the parallel component $B_{\parallel l}$, and a systematic increase of the out-of-plane component $B_z$.
The electric field increases systematically in all components.
It is interesting that the component $E_z$ remains very uniform in the process.
When we subtracted the ideal field contribution $E_z' = E_z - (\left<\bm\beta\right>\times\bm{B})_z$, we found that $E_z' \simeq E_{\parallel B}$.
The non-ideal field component peaks in the middle of the layer, and has thickness scale comparable to the $v_z$ structure.

\begin{figure*}
\centering
\includegraphics[width=\textwidth]{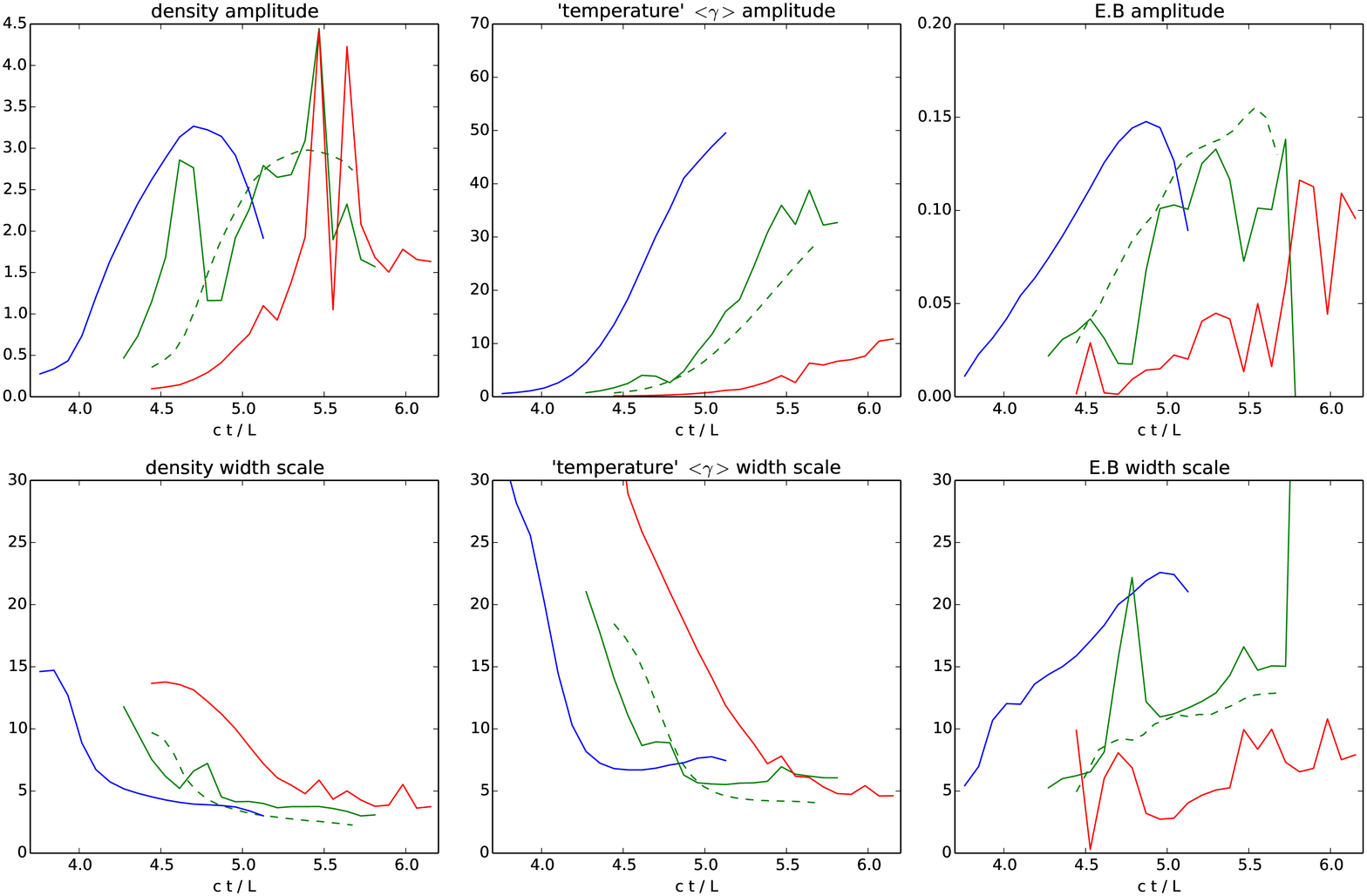}
\caption{Evolution of the transverse structure of the current layer in the linear phase for runs: s14L1600 (\emph{red lines}), s27L1600 (\emph{solid green lines}), s27L800 (\emph{dashed green lines}), and s55L1600 (\emph{blue lines}). We have calculated the transverse profiles of particle number density (\emph{left panels}), average particle energy (\emph{middle panels}) and non-ideal electric field (\emph{right panels}) across the centroid of the current layer. These profiles were fitted with a model consisting of a Gaussian and a constant background. \emph{The upper panels} show the evolution of the amplitude of the Gaussian model, and \emph{the lower panels} show the evolution of the width scale in units of $\rho_0$. Occasional sharp spikes are due to the passage of plasmoids.}
\label{fig_current_stats}
\end{figure*}

We fitted the perpendicular profiles of particle number density $n$, average particle energy $\left<\gamma\right>$, and $\bm{E}\cdot\bm{B}$ with Gaussian models with uniform background.
In Figure \ref{fig_current_stats}, we show the evolution of the amplitudes and perpendicular dispersions of the three parameters.
This reveals the complex structure of the current layer, with the three parameters characterized by different thickness scales. The general trend in time is an increase in the amplitudes of all three parameters, the thinning of the density and average energy profiles, and the broadening of the $\bm{E}\cdot\bm{B}$ profile.
The amplitudes of the average particle energy grow exponentially in time as long as we can measure their profile, with the growth rate increasing systematically with the magnetization $\sigma_{\rm hot}$.
On the other hand, the amplitudes of $\bm{E}\cdot\bm{B}$ appear to grow linearly in time, and the growth rate is similar for $\sigma_{\rm hot} = 2.7,5.5$.
The width scales of the density and temperature profiles converge to a value $\sim 5\rho_0$, roughly independent of $\sigma_{\rm hot}$.
The difference between the width scales measured for the density and temperature profiles is probably not significant.
However, the width scale of the $\bm{E}\cdot\bm{B}$ clearly depends on and increases with $\sigma_{\rm hot}$.
We suggest that the width scale of the density and temperature profiles corresponds roughly to the plasma skin depth $d_{\rm e}$, and that the width scale of the $\bm{E}\cdot\bm{B}$ region corresponds to the typical gyroradius of particles heated in the layer $\rho \simeq \left<\gamma\right>\rho_0$.

\begin{figure}
\centering
\includegraphics[width=\columnwidth]{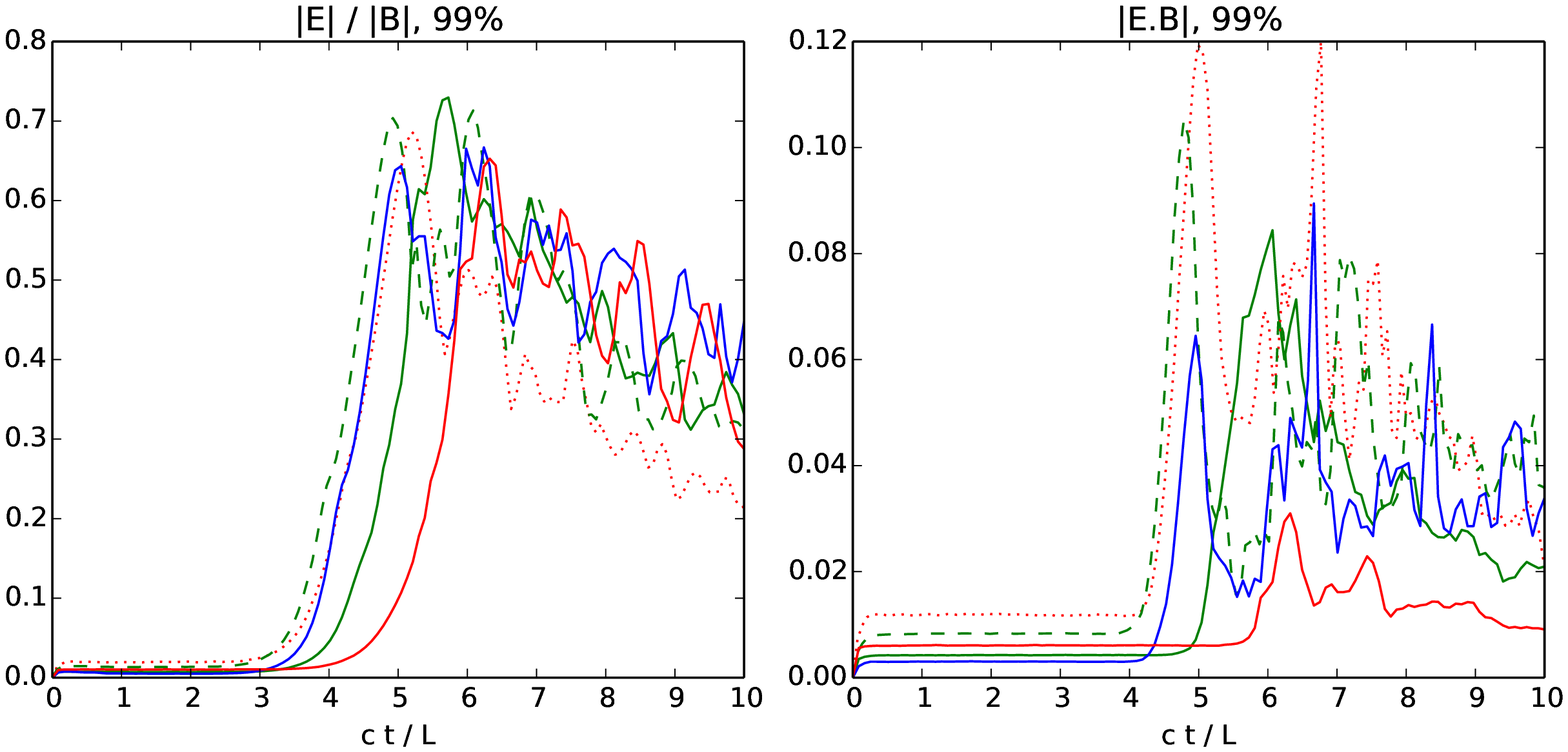}
\caption{Evolution of the 99 percentile volume-weighted values of $|\bm{E}|/|\bm{B}|$ (\emph{left panel}) and $|\bm{E}\cdot\bm{B}|/B_0^2$ (\emph{right panel}) compared for several runs. Line colors and types are the same as in Figure \ref{fig_total_energy}.}
\label{fig_hist_EB}
\end{figure}

We also calculated the volume-weighted distribution functions for key scalar quantities $|\bm{E}|/|\bm{B}|$ and $|\bm{E}\cdot\bm{B}|$.
Figure \ref{fig_hist_EB} shows the 99 percentile values as functions of simulation time.
The values of $|\bm{E}|/|\bm{B}|$ consistently reach the level of $\simeq 0.7$, independently of the setup parameters, at about the time of linear instability saturation.
The situation with the $|\bm{E}\cdot\bm{B}|$ values is less clear, with a number of sharp peaks observed at different times.
The smallest values $|\bm{E}\cdot\bm{B}| \sim 0.03B_0^2$ are recorded for Run s14L1600, which is qualitatively consistent with the results shown in Figure \ref{fig_current_stats}.
For other runs, the peak values are in the range $|\bm{E}\cdot\bm{B}| \sim (0.08-0.12)B_0^2$.

\subsection{Particle acceleration - total spectra}

\begin{figure*}
\centering
\includegraphics[width=0.49\textwidth]{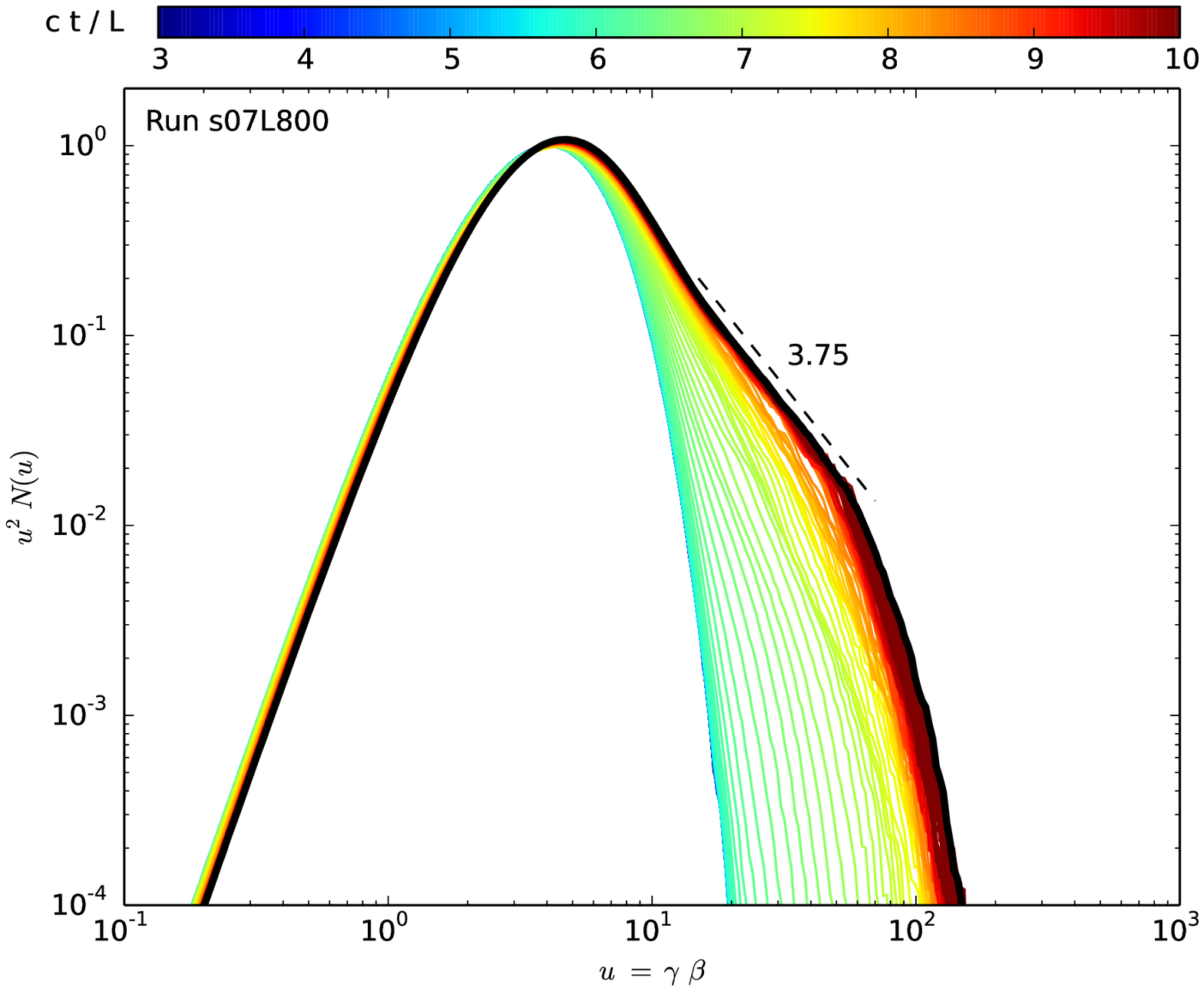}
\includegraphics[width=0.49\textwidth]{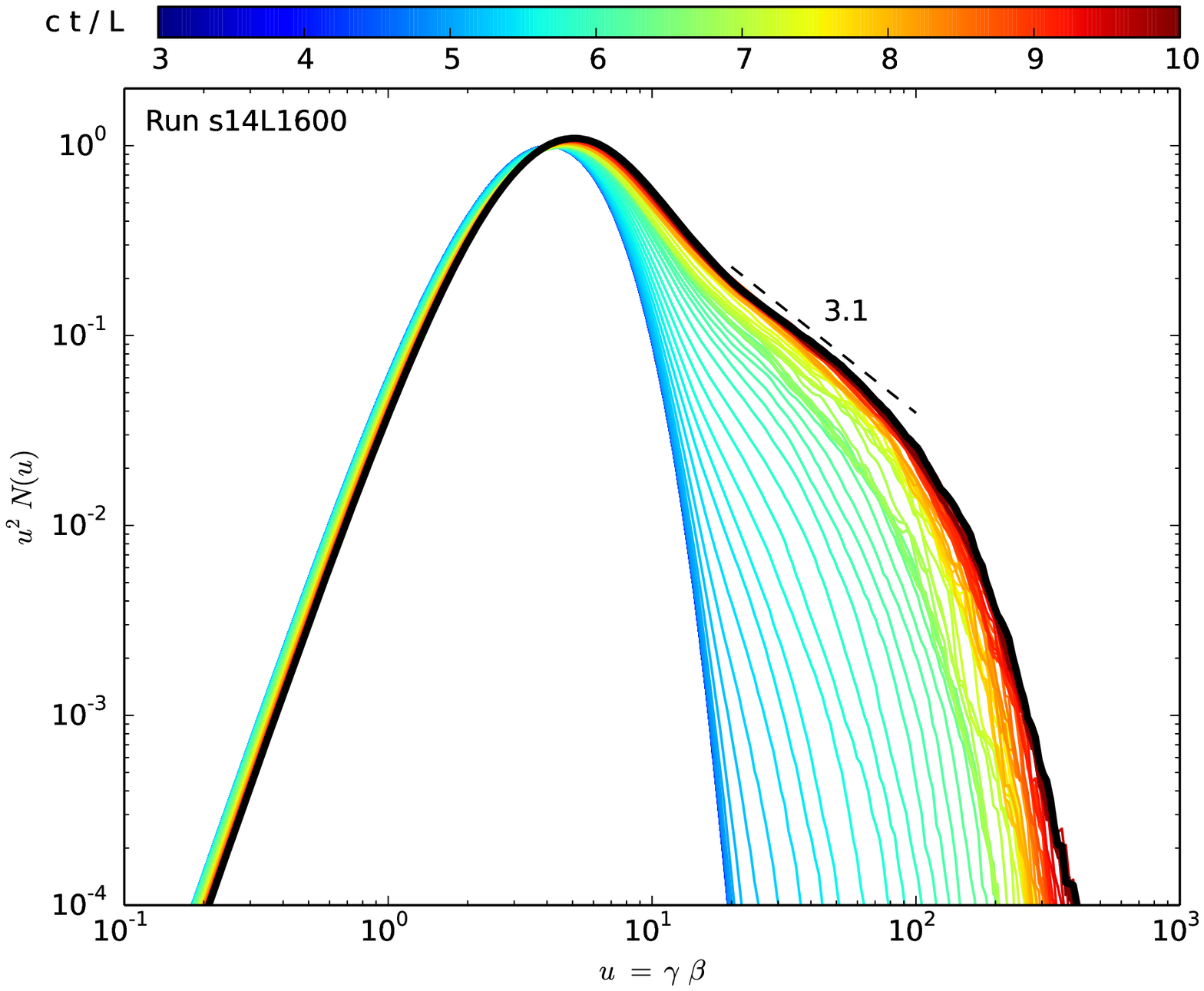}
\includegraphics[width=0.49\textwidth]{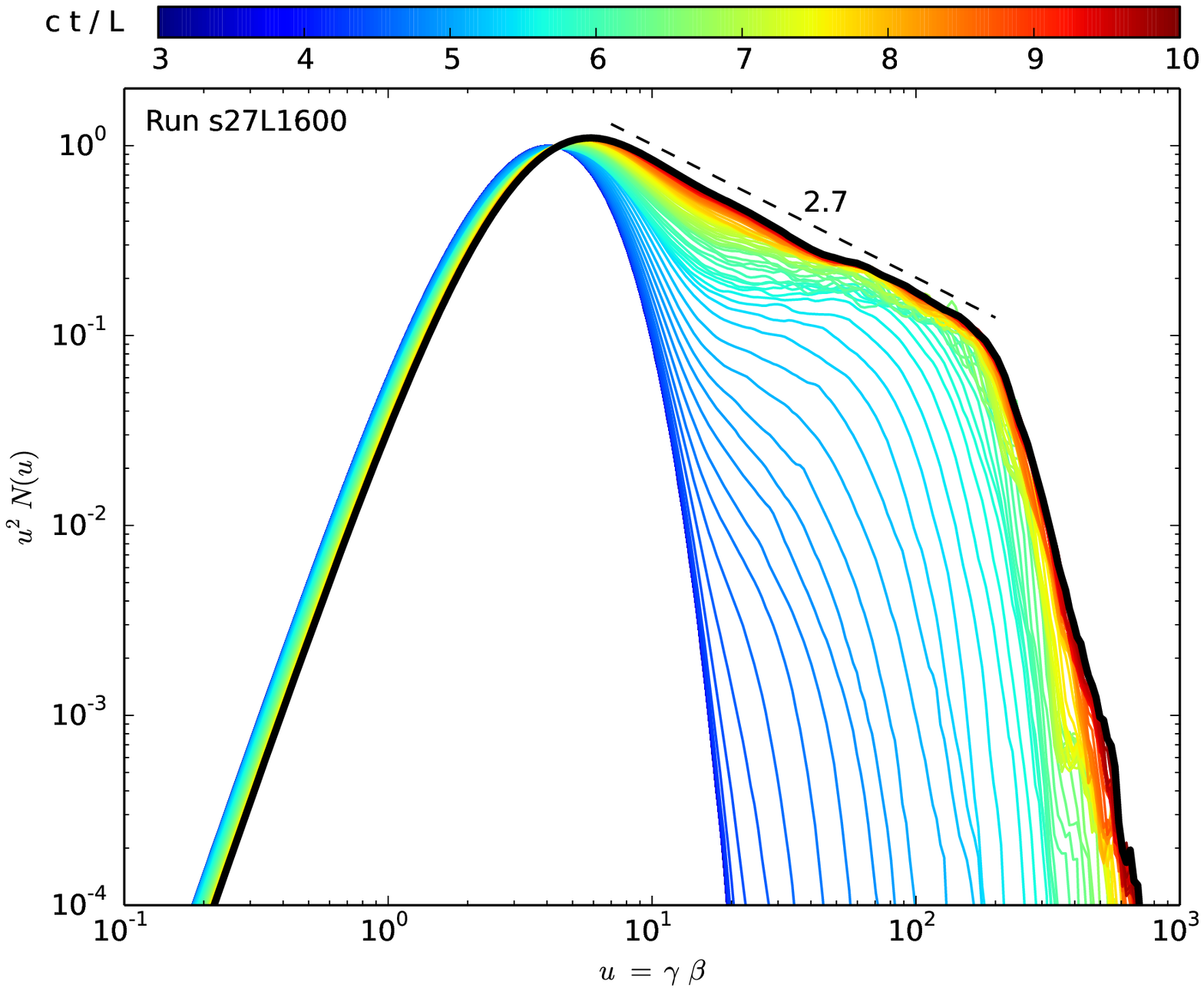}
\includegraphics[width=0.49\textwidth]{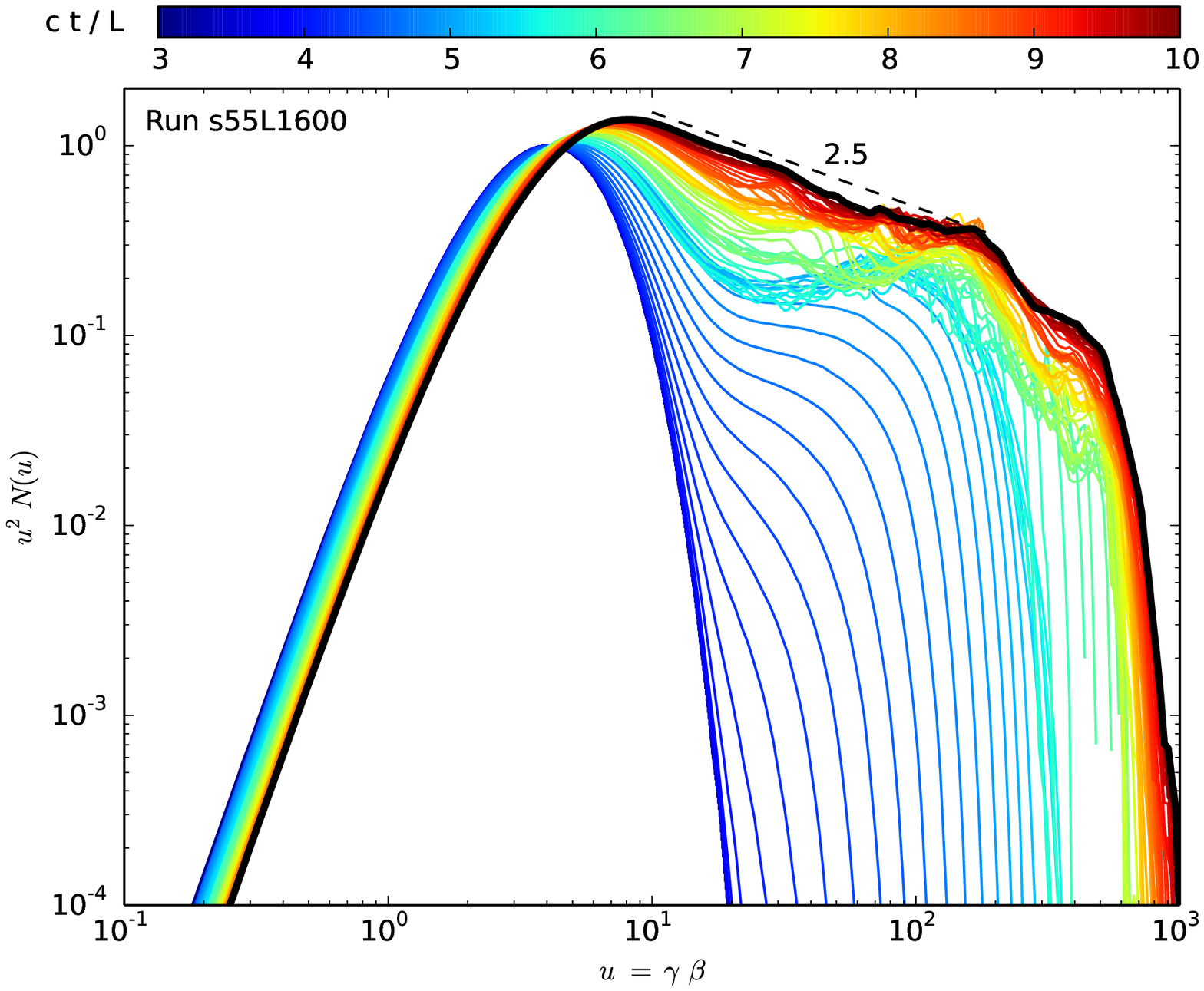}
\caption{Evolution of the momentum distributions $u^2N(u)$ of electrons and positrons for selected runs. The color scale indicates simulation time normalized to the light-crossing time scale $L/c$. The distributions are sampled at linearly uniform time intervals, however, the intervals are different for each simulation.}
\label{fig_spectra}
\end{figure*}

In Figure \ref{fig_spectra}, we show the momentum distributions $u^2 N(u)$ of electrons and positrons as functions of simulation time. The distributions are normalized to the peak of the initial distribution, which in all cases is the Maxwell-J\"{u}ttner distribution for $\Theta = 1$. The initial distribution begins to evolve at $ct/L \simeq 4$ (blue) in a period of rapid and regular (linear) energization of a relatively small fraction of all particles. The uniformity of the individual distributions on the log-log plot suggests an exponential growth of both the maximum particle energy and the number fraction of participating particles. For the case of Run s55L1600, this acceleration phase is efficient enough to form a second peak of the momentum distribution. The linear acceleration phase eventually ceases, and the subsequent non-linear evolution of the momentum distribution leads to a gradual formation of a power-law. The ``valley'' between the initial Maxwellian component and the high-energy bump is promptly leveled. The maximum particle energy increases slowly but systematically in several steps corresponding to the global oscillation of the merged magnetic domains.

We have roughly estimated the index $p$ of the power-law tail of the distribution $N(u) \propto u^{-p}$ for each simulation at $ct/L \sim 10$. The values are reported for selected runs in Figure \ref{fig_spectra}. There is a clear trend of $p$ increasing with $\sigma_{\rm hot}$, from $p \simeq 3.5$ for $\sigma_{\rm hot} = 0.7$ to $p \simeq 2.5$ for $\sigma_{\rm hot} = 5.5$. We have not attempted to measure the power-law index rigorously as a function of the simulation time, as accurate automated decomposition of the distribution function requires very complicated modeling \citep{Wer16}.

\begin{figure*}
\centering
\includegraphics[width=\textwidth]{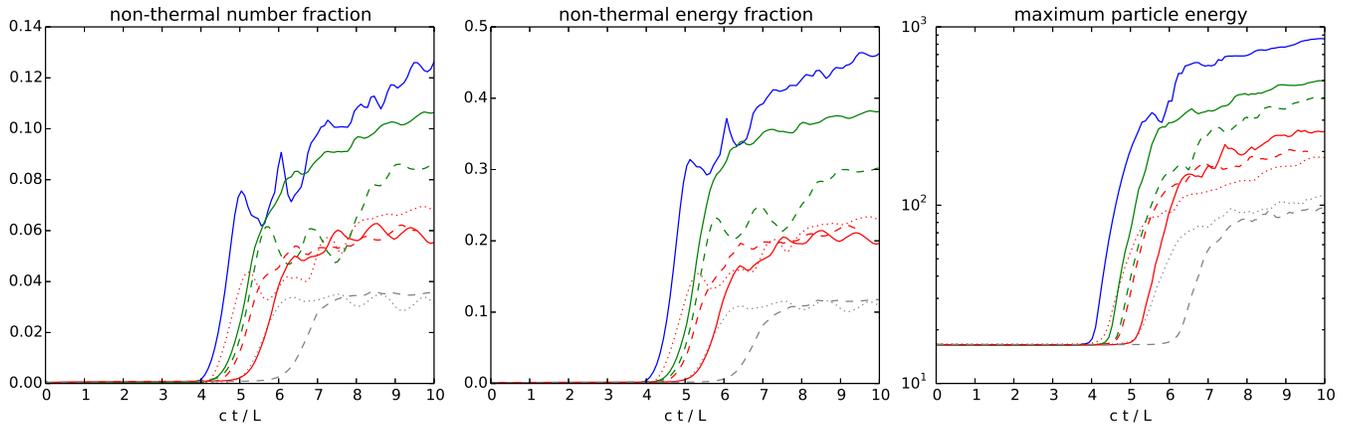}
\caption{
\emph{Upper left panel:} time evolution of the number fraction contained in the non-thermal high-energy tail of the electron distribution.
\emph{Upper right panel:} time evolution of the energy fraction contained in the high-energy tail.
\emph{Lower left panel:} time evolution of the maximum particle Lorentz factor measured at the level of $u^2N(u) = 10^{-3}$ for normalized electron distributions.
Line colors and types are the same as in Figure \ref{fig_total_energy}.
}
\label{fig_nonthermal_fraction}
\end{figure*}

Instead, we investigate what is the fraction of all particles contained in the high-energy tail by subtracting the low-energy Maxwellian component. In Figure \ref{fig_nonthermal_fraction}, we plot as functions of simulation time: the fraction of the particle number $f_{\rm n}$, the fraction of the particle energy $f_{\rm e}$, and the maximum particle energy $\gamma_{\rm max}$, defined formally at the level of $u^2 N(u) = 10^{-3}$ for the normalized distributions.
During the linear acceleration phase, both $f_{\rm n}$, $f_{\rm e}$ and $\gamma_{\rm max}$ grow roughly exponentially with time.
In the non-linear phase, the growth becomes much slower, although it does not fully saturate before $ct/L = 10$, especially for higher values of $\sigma_{\rm hot}$.

\begin{figure}
\centering
\includegraphics[width=\columnwidth]{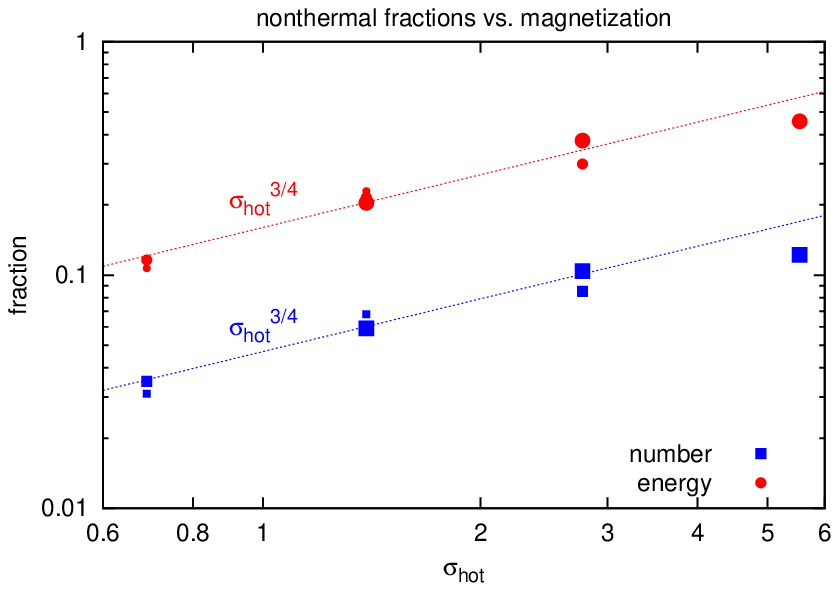}
\includegraphics[width=\columnwidth]{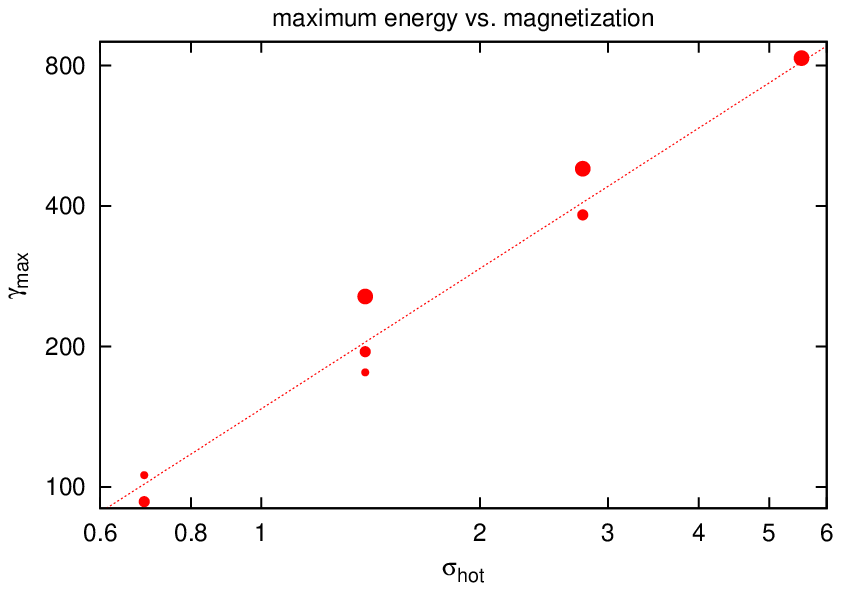}
\caption{
\emph{Left panel:} dependence of the number (blue squares) and energy (red circles) fractions contained in the high-energy tail of the electron distribution, averaged for $ct/L \in [9:10]$, on the mean hot magnetization value.
\emph{Right panel:} dependence of the maximum Lorentz factor measured at the level of $u^2N(u) = 10^{-3}$ for normalized electron distributions, averaged for $ct/L \in [9:10]$.
The point size indicates the simulation domain size $L$.}
\label{fig_nonthermal_fraction2}
\end{figure}

The values of $f_{\rm n}$, $f_{\rm e}$ and $\gamma_{\rm max}$ at $ct/L \simeq 10$ for all simulations are reported in Table \ref{table_runs} and plotted in Figure \ref{fig_nonthermal_fraction2}.
The number fraction ranges from $f_{\rm n} \simeq 0.03$ for $\sigma_{\rm hot} = 0.7$ to $f_{\rm n} \simeq 0.17$ for $\sigma_{\rm hot} = 5.5$.
The corresponding values of the energy fraction are $f_{\rm e} \simeq 0.11$ and $f_{\rm e} \simeq 0.59$.
Both $f_{\rm n}$ and $f_{\rm e}$ appear to scale with the magnetization like $\sigma^{3/4}$, with the exception of the highest $\sigma$ value,
although simulations with even greater $\sigma$ are needed to verify this scaling.
In any case, this trend could not continue to very high values of $\sigma$, as $f_{\rm n},f_{\rm e} < 1$.
The ratio $f_{\rm e}/f_{\rm n}$, corresponding to the ratio of average particle energies $\left<\gamma\right>_{\rm nth}/\left<\gamma\right>_{\rm tot}$, shows a slight increase from $f_{\rm e}/f_{\rm n} \simeq 3.3$ for $\sigma_{\rm hot} = 0.7$ to $f_{\rm e}/f_{\rm n} \simeq 3.7$ for $\sigma_{\rm hot} = 5.5$.
The maximum particle energy ranges from $\gamma_{\rm max} \simeq 100$ for $\sigma_{\rm hot} = 0.7$ to $\gamma_{\rm max} \simeq 800$ for $\sigma_{\rm hot} = 5.5$, and is consistent with a linear dependence on $\sigma$.

\subsection{Individual particle histories}

\begin{figure*}
\centering
\includegraphics[width=\textwidth]{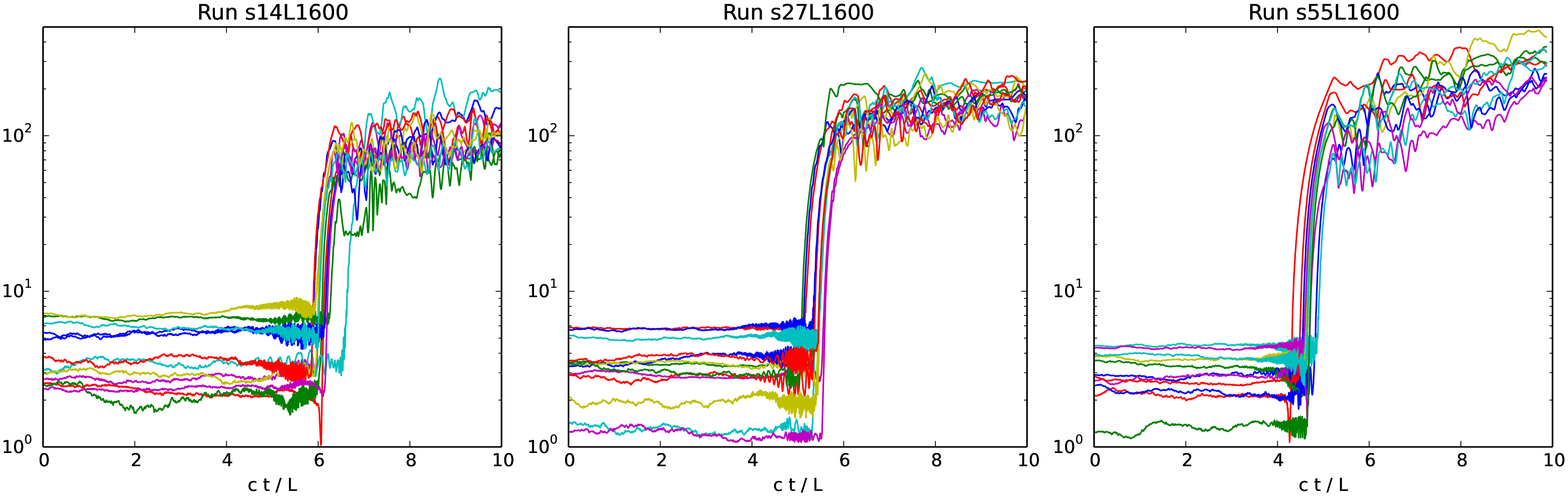}
\caption{Energy histories $\gamma(t)$ for individual particles selected for their high energy at $ct/L = 10$. Each diagram shows roughly $\sim 10$ most energetic particles.}
\label{fig_orbit_sample}
\end{figure*}

At the beginning of each simulation, we randomly selected $10^5$ individual particles in order to record their detailed histories. 
We will now use this information in order to better understand particle acceleration in the linear stage of the simulation.
In Figure \ref{fig_orbit_sample}, we show a sample of particle energy histories $\gamma(t)$ for particles that obtain the highest energies at $ct/L = 10$.
For every particle, this reveals a sharp acceleration episode during the saturation of the linear instability.
Now we ask the question, whether the difference in the final energy values obtained by these most energetic particles for each simulation is due to stronger effective value of the parallel electric field, or rather due to longer acceleration time scale.
To this end, we have analyzed the energy histories $\gamma(t)$, identifying episodes $[t_1:t_2]$ of the largest monotonic energy gain $\Delta\gamma = \gamma(t_2) - \gamma(t_1)$.
We applied a mild smoothing to $\gamma(t)$, and we excluded episodes with $\gamma(t_1) \gtrsim 0.2\gamma_{\rm max}$.
Then, we calculated the mean electric field accelerating the particle from $\left<E\right>/B_0 = \Delta\gamma/(\omega_0\Delta t)$, where $\Delta t = t_2-t_1$ and $\omega_0 = c/\rho_0$.

\begin{figure*}
\centering
\includegraphics[width=\textwidth]{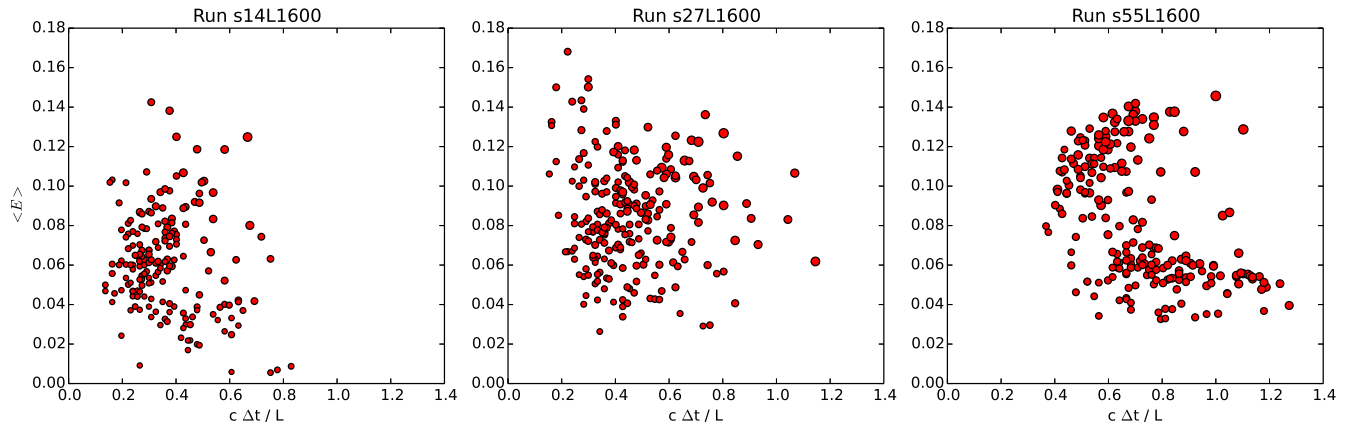}
\caption{Duration vs. mean electric field calculated for the main acceleration episodes for representative samples of energetic individually tracked particles. The area of the markers is proportional to the final particle energy $\gamma_{\rm fin}$ calculated for $ct/L = 10$. Each diagram shows roughly $\sim 200$ most energetic particles.}
\label{fig_orbit_accel}
\end{figure*}

In Figure \ref{fig_orbit_accel}, we show the distribution of $\left<E\right>/B_0$ vs. $c\Delta t/L$ for three simulations with different magnetization values, indicating also the final particle energy $\gamma_{\rm fin}$ at $ct/L = 10$.
For each simulation, there is a considerable spread in the values of $\left<E\right>$ and $\Delta t$ calculated for individual energetic particles.
While $\left<E\right>$ and $\Delta t$ determine the energy gain $\Delta\gamma$ for the main acceleration episode, there is also a substantial scatter between $\Delta\gamma$ and $\gamma_{\rm fin}$.
The mean values of $\left<E\right>/B_0$ are $0.053, 0.071,0.067$ for $\sigma_{\rm hot} = 1.4,2.7,5.5$, respectively, and the corresponding mean values of $c\Delta t/L$ are $0.32,0.39,0.59$. With this, we can account for a factor of $\simeq 2.4$ difference between the maximum particle energy between $\sigma_{\rm hot} = 1.4$ and $\sigma_{\rm hot} = 5.5$, while according to Table \ref{table_runs} the difference between $\gamma_{\rm max}$ values for these simulations is by factor $3.1$.
The difference appears to be both due to increase in typical electric field strength between $\sigma_{\rm hot} = 1.4$ and $\sigma_{\rm hot} = 2.7$, as well as due to longer acceleration time scale between $\sigma_{\rm hot} = 2.7$ and $\sigma_{\rm hot} = 5.5$.

\section{Discussion}
\label{sec_dis}

Simulating force-free magnetostatic equilibria offers a new approach for studying relativistic magnetic dissipation, an alternative to simulating the Harris-type current layers.
A major conceptual advantage of this approach is that there are no current layers or any structures on the kinetic scale in the initial plasma configuration.
The formation of current layers results from the saturation of an ideal plasma instability, and this rapidly creates conditions for efficient particle acceleration.
The most energetic particles follow the evolving current layers until they disintegrate on the dynamical time scale, and afterwards they interact chaotically with the slowly decaying turbulence.

The initial configuration consists of two positive and two negative domains of magnetic vector potential $A_z \propto B_z$.
These symmetrically interlocked domains attract each other and hence they remain in an unstable equilibrium until the particle distribution noise breaks the positional symmetry.
This triggers an exponential increase in the bulk plasma velocities $\bm\beta$, and hence an exponential increase in the ideal electric fields $\bm{E} = \bm{B}\times\bm\beta$.
Figure \ref{fig_total_energy} shows that this instability is not seen in the time evolution of the global non-ideal electric energy.
The e-folding growth rate $\tau$ of the total electric energy is measured to scale with the mean hot magnetization $\sigma_{\rm hot}$ roughly like $c\tau(\sigma_{\rm hot})/L \simeq 0.13/v_{\rm A}(0.21\sigma_{\rm hot})$.
In the limit of $\sigma_{\rm hot} \gg 1$, this is in excellent agreement with the value $c\tau/L \simeq 0.127$ (see Figure \ref{fig_growth_rates}) measured in the FF simulations \citep{Eas15}.

The origin of the current layers can be traced to the converging motion of two magnetic domains resulting in a pile-up of particles at the magnetic null points.
The exponential increase in the plasma velocity creates an exponentially increasing flux of particles forced into the current layers.
This driven reconnection is in contrast to the undriven reconnection resulting from relaxed Harris-type current layers, where magnetic diffusion regions are characterized by low plasma density.
The current layers in our simulations systematically grow in length, they become hotter and thinner, and they are subject to only mild tearing mode instability.
We observe that around the saturation moment of the ideal instability the inflow of plasma into the current layer is interrupted, and the current layer is rapidly disintegrated by internal motions of the newly formed magnetic domain encompassing the layer.

Throughout the evolution of the current layers, we observe a systematic increase and smooth structure across the layer of the total electric field.
This field originated outside the current layers from the motions of ideal plasma as $\bm{E} = \bm{B}\times\bm\beta$.
In the ideal MHD limit, such field should vanish in the middle of the current layer, and we would strictly expect that $\bm{E}\cdot\bm{B} = 0$.
The ideal MHD condition is obviously violated inside the current layer, and $\bm{E}\cdot\bm{B}$ is dominated by $E_zB_z$.
The smooth structure of the $E_z$ component suggests that it is advected into the layer from outside, rather than created locally in a separate process.
The thickness scale of the $\bm{E}\cdot\bm{B} \ne 0$ region is significantly larger than the thickness scale of the particle density pile-up, moreover, it is systematically increasing with the simulation time.
This additional thickness scale is roughly of the order of the gyroradius corresponding to the mean particle energy inside the layer, and it can be seen imprinted into the profile of the current density.
The Vlasov momentum equation dictates that for each particle species the following must be satisfied:
$qn(\bm{E}\cdot\bm{B}) = B_i(\partial{U_i}/\partial{t} + \partial{P_{ij}}/\partial{x_j})$,
where $\bm{U} = mn\left<\gamma\bm{v}\right>$ is the momentum density, and $P_{ij} = mn\left<\gamma v_iv_j\right>$ is the pressure tensor.
We have checked that in general the pressure gradients do not support the $\bm{E}\cdot\bm{B}$, and hence the particle momentum density has to increase, i.e., particles are collectively heated in the $\bm{E}\cdot\bm{B} \ne 0$ regions.

The particle energy distributions evolve from the initial relativistic Maxwell-J\"{u}ttner to form a high-energy power-law tail on the time scale of $\sim 10L/c$ in two stages.
In the first, linear stage, particles are heated by regular electric fields in the linearly growing current layers.
These particles produce a distinct high-energy bump in the energy distribution, with the number of heated particles and their maximum energy increasing exponentially with time.
In the second, non-linear stage, particles energized in the linear stage become detached from the magnetic potential, and they interact chaotically with secondary current layers arising in the slowly damped turbulent plasma motions, diffusing in the energy space.
The power-law distributions $N(u)\propto u^{-p}$ arise in the non-linear stage, however, their indices appear to be determined already in the linear stage.
They can be robustly predicted by the ratio of the two distribution peaks at the saturation moment of the linear instability, or in other words, by the energy fraction of particles interacting with the current layers in the linear stage.

We showed that the acceleration process in the linear stage is very regular with both the maximum particle energy and the number fraction of the high-energy particles increasing exponentially with time.
On the other hand, we showed that individual particles are accelerated linearly in time, as if subjected to a roughly constant value of $E_{\parallel v}$.
The exponential growth of the number fraction of high-energy particles can be related to the exponential growth of the flux of particles joining the current layer.
The exponential growth of the maximum particle energy $\gamma_{\rm max}$ in the linear stage is consistent with the growth of the peak value of the non-ideal electric field component $E_{\parallel B}$.
In the non-linear stage, the growth of $\gamma_{\rm max}$ is close to linear, which would correspond to a constant value of $E_{\parallel B}$.
The final values of the non-thermal particle fractions appear to scale with the mean magnetization as $f_{\rm n,e} \propto \sigma_{\rm hot}^{3/4}$.
These scalings are expected to saturate below unity for $\sigma_{\rm hot} \gg 1$, and the $3/4$ index may result from a transition between non-relativistic and ultra-relativistic regimes.
On the other hand, the maximum particle energy scales roughly like $\gamma_{\rm max} \propto \sigma_{\rm hot}$, as has been found for sufficiently large simulations of relativistic reconnection initiated from the Harris-type layers \citep{Wer16}.
Figure \ref{fig_orbit_accel} shows that the final particle energy is determined both by the non-ideal electric field strength and by the time spent by the particle in the acceleration region.

The emergence of the power-law component can be attributed to the second-order Fermi process.
This process can be understood in the following way --- (1) high-energy particles emerging from the linear current layers are characterized by large gyroradii, and they follow on chaotic orbits, independently of the small-scale magnetic structures; (2) oscillations of the merged magnetic structures create multiple short-lived current layers with different orientations of the non-ideal electric field; (3) high-energy particles are more likely to pass across these new acceleration regions, however they have comparable chances to be accelerated or decelerated, depending on the sign of $E_{\parallel v}$.
We do not find any clear signatures of the first-order Fermi process in the non-linear acceleration phase \citep[cf.][]{Hos12,Ber16}.

The power-law indices for the highest mean cold magnetization probed $\sigma_{\rm cold} \sim 30$ are $p \simeq 2.5$, which is significantly higher than the values $p \sim 1.2 - 1.7$ reported for Harris-layer simulations with comparable upstream values of $\sigma_{\rm cold}$ \citep{Sir14,Guo14,Wer16}.
This may actually be useful for astrophysical applications, where there are more cases for high particle distribution indices $p > 2$; in this sense the Harris-type reconnection in the ultra-relativistic limit appears to be too efficient a particle accelerator.
The lower acceleration efficiency can be explained by the lower fraction of magnetic free energy --- $f_{\rm B} \simeq 25\%$ in our case vs. $f_{\rm B} \simeq 50\%$ in the case of Harris-type layers.
The fraction of free magnetic energy in our 2D simulations is only slightly lower than the theoretical prediction of $29\%$ corresponding to the isotropic Taylor state (see Section \ref{sec_setup}).
In the case of periodic Harris-type setups, the final configuration departures further from the Taylor state, and hence an analytical prediction of $f_{\rm B}$ is more difficult.
Finally, we note that the free energy associated with ABC magnetic fields can be increased indefinitely by considering higher-order harmonics.
However, in such case the topological constraints limit the dissipation efficiency in 2D, and hence investigating highly efficient magnetic dissipation requires large 3D simulations \citep{ZraEas16}.

A certain disadvantage of the ``ABC'' magnetostatic equilibria is that their mean magnetization $\sigma$ is numerically limited by the spatial resolution $L/\rho_0$, at least under our conservative requirement that the kinetic scale $\rho_0$ should always be resolved.
This limitation arises from the uniform distribution of magnetic field gradients that have to be supported by volume-filling currents.
Effectively, our study is confined to one order of magnitude in plasma magnetization. 
This is in contrast to the Harris-type configuration, where magnetic field gradients are confined to narrow layers, and hence a wide range of upstream magnetizations, spanning almost three orders of magnitude, can be studied at fixed numerical resolution \citep{Guo14,Wer16}.

Our simulations correspond effectively to guide-field reconnection, and given the Beltrami condition we have no way to vary the level of the guide field.
Even though $B_z$ vanishes initially along the separatrices, it is readily advected, roughly incompressibly, into the current layers, so that $|\bm{E}| < |\bm{B}|$ at all times everywhere in the layer (see Figure \ref{fig_hist_EB}).
This is a potential concern if we would like to apply these results to the gamma-ray flares in the Crab Nebula, as the prospect for breaking the synchrotron photon energy limit in the Harris-layer reconnection scenario depends on the guide field \citep{Cer13,Cer14}.
However, what we need in order to break the synchrotron limit is $E_{\parallel v} > B_{\perp v}$, and these components depend on the exact trajectories of energetic particles.
Simulations of force-free magnetostatic equilibria taking into account the synchrotron radiation reaction are required to address this problem, and will be presented in a future study (Y.~Yuan et~al., in preparation).

\section{Conclusions}
\label{sec_con}

We have investigated magnetic dissipation and particle acceleration in the simplest case of relativistic unstable 2D ``ABC'' magnetostatic equilibria by means of fully kinetic PIC simulations with pair plasma.
An ideal instability leads to the formation of dynamical current layers with complex internal structure, where a fraction of particles is heated collectively by non-ideal electric fields.
Saturation of the instability leads to a slowly decaying turbulence which diffuses the energetic particles in the energy space, effectively forming power-law energy distributions.
The power-law slopes are significantly softer ($p \gtrsim 2.5$) when compared with the case of ultra-relativistic reconnection initiated from the Harris-type layers ($p \lesssim 1.5$).
This kind of magnetic dissipation corresponds to driven relativistic reconnection with a substantial guide field.
However, since most of the dissipation and particle acceleration proceed over a single light-crossing time scale, this is an attractive scenario for the production of rapid high-energy radiation flares observed in blazars, pulsar wind nebulae, GRBs, etc.

\acknowledgments 

We thank Maxim Lyutikov and Lorenzo Sironi for stimulating discussions.
K.N. was supported by NASA through Einstein Postdoctoral Fellowship grant number PF3-140130 awarded by the Chandra X-ray Center, which is operated by the Smithsonian Astrophysical Observatory for NASA under contract NAS8-03060.
We acknowledge the use of computational resources obtained from XSEDE (Stampede), University of Colorado Research Computing (Janus), and KIPAC/SLAC (Bullet).



\begin{thebibliography}{}

\bibitem[{Abdo} {et~al.}(2009)]{Abd09}
{Abdo}, A.~A., {et~al.}, 2009, Nature, 462, 331

\bibitem[{Abdo} {et~al.}(2011)]{Abd11}
{Abdo}, A.~A., {Ackermann}, M., {Ajello}, M., {et~al.}, 2011, Science, 331, 739

\bibitem[{Aharonian} {et~al.}(2007)]{Aha07}
{Aharonian}, {F.}, {et~al.}, 2007, ApJ, 664, L71

\bibitem[{Aleksi{\'c}} {et~al.}(2011)]{Ale11}
{Aleksi{\'c}}, J., {et~al.}, 2011, ApJ, 730, L8

\bibitem[Aleksi{\'c} et~al.(2014)]{Ale14}
Aleksi{\'c}, J., Ansoldi, S., Antonelli, L.~A., et~al.\ 2014, Science, 346, 1080

\bibitem[Baty et~al.(2013)]{Bat13}
Baty, H., Petri, J., \& Zenitani, S.\ 2013, MNRAS, 436, L20

\bibitem[Begelman(1998)]{Beg98}
Begelman, M.~C., 1998, ApJ, 493, 291

\bibitem[{Begelman} {et~al.}(2008)]{Beg08}
{Begelman}, M.~C., {Fabian}, A.~C., {Rees}, M.~J., 2008, MNRAS, 384, L19

\bibitem[Beresnyak \& Li(2016)]{Ber16}
Beresnyak, A., \& Li, H.\ 2016, ApJ, 819, 90

\bibitem[Blandford et~al.(2014)]{Bla14}
Blandford, R., Simeon, P., \& Yuan, Y.\ 2014, NuPhS, 256, 9

\bibitem[Blandford et~al.(2015)]{Bla15}
Blandford, R., East, W., Nalewajko, K., Yuan, Y., \& Zrake, J.\ 2015, arXiv:1511.07515

\bibitem[Brandenburg {et~al.}(2015)]{Brandenburg2015}
Brandenburg, A., Kahniashvili, T., \& Tevzadze, A.~G. 2015, PhRvL, 114, 075001

\bibitem[Buehler et~al.(2012)]{Bue12}
Buehler, R., Scargle, J.~D., Blandford, R.~D., et~al.\ 2012, ApJ, 749, 26

\bibitem[Cerutti et~al.(2013)]{Cer13}
Cerutti, B., Werner, G.~R., Uzdensky, D.~A., \& Begelman, M.~C.\ 2013, ApJ, 770, 147

\bibitem[Cerutti et~al.(2014)]{Cer14}
Cerutti, B., Werner, G.~R., Uzdensky, D.~A., \& Begelman, M.~C.\ 2014, ApJ, 782, 104

\bibitem[Clausen-Brown \& Lyutikov(2012)]{Cla12}
Clausen-Brown, E., \& Lyutikov, M.\ 2012, MNRAS, 426, 1374

\bibitem[Coroniti(1990)]{Cor90}
Coroniti, F.~V.\ 1990, ApJ, 349, 538

\bibitem[East et~al.(2015)]{Eas15}
East, W.~E., Zrake, J., Yuan, Y., \& Blandford, R.~D.\ 2015, PhRvL, 115, 095002

\bibitem[{Giannios} {et~al.}(2009)]{Gia09}
{Giannios}, {D.}, {Uzdensky}, {D.~A.}, \& {Begelman}, {M.~C.}, 2009, MNRAS, 395, L29

\bibitem[Guo et~al.(2014)]{Guo14}
Guo, F., Li, H., Daughton, W., \& Liu, Y.-H.\ 2014, PhRvL, 113, 155005

\bibitem[Hayashida et~al.(2015)]{Hay15}
Hayashida, M., Nalewajko, K., Madejski, G.~M., et~al.\ 2015, ApJ, 807, 79

\bibitem[Hoshino(2012)]{Hos12}
Hoshino, M.\ 2012, PhRvL, 108, 135003

\bibitem[Kirk \& Skj{\ae}raasen(2003)]{Kir03}
Kirk, J.~G., \& Skj{\ae}raasen, O.\ 2003, ApJ, 591, 366

\bibitem[Liu et~al.(2015)]{Liu15}
Liu, Y.-H., Guo, F., Daughton, W., Li, H., \& Hesse, M.\ 2015, PhRvL, 114, 095002

\bibitem[{Lovelace} {et~al.}(1997)]{Lov97}
{Lovelace}, R.~V.~E., {Newman}, W.~I., \& {Romanova}, M.~M., 1997, ApJ, 484, 628

\bibitem[Lyubarsky \& Kirk(2001)]{Lyu01}
Lyubarsky, Y., \& Kirk, J.~G.\ 2001, ApJ, 547, 437

\bibitem[Mizuno et~al.(2012)]{Miz12}
Mizuno, Y., Lyubarsky, Y., Nishikawa, K.-I., \& Hardee, P.~E.\ 2012, ApJ, 757, 16

\bibitem[{Nalewajko} {et~al.}(2011)]{Nal11}
{Nalewajko}, K., {Giannios}, D., {Begelman}, M.~C., {Uzdensky}, D.~A., \& {Sikora}, M., 2011, MNRAS, 413, 333

\bibitem[{O'Neill} {et~al.}(2012)]{ONe12}
{O'Neill}, S.~M., {Beckwith}, K., \& {Begelman}, M.~C., 2012, MNRAS, 422, 1436

\bibitem[Saito et~al.(2013)]{Sai13}
Saito, S., Stawarz, {\L}., Tanaka, Y.~T., et~al.\ 2013, ApJ, 766, L11

\bibitem[Sironi \& Spitkovsky(2011)]{Sir11}
Sironi, L., \& Spitkovsky, A.\ 2011, ApJ, 741, 39

\bibitem[Sironi \& Spitkovsky(2014)]{Sir14}
Sironi, L., \& Spitkovsky, A.\ 2014, ApJ, 783, L21

\bibitem[Syrovatskii(1966)]{Syr66}
Syrovatskii, S.~I.\ 1966, \sovast, 10, 270

\bibitem[Tavani et~al.(2011)]{Tav11}
Tavani, M., Bulgarelli, A., Vittorini, V., et~al.\ 2011, Science, 331, 736

\bibitem[{Taylor(1974)}]{Tay74}
Taylor, J.~B. 1974, Physical Review Letters, 33, 1139

\bibitem[Uzdensky et~al.(2011)]{Uzd11}
Uzdensky, D.~A., Cerutti, B., \& Begelman, M.~C.\ 2011, ApJ, 737, L40

\bibitem[Werner et~al.(2016)]{Wer16}
Werner, G.~R., Uzdensky, D.~A., Cerutti, B., Nalewajko, K., \& Begelman, M.~C.\ 2016, ApJ, 816, L8

\bibitem[Yee(1966)]{Yee66}
Yee, K.\ 1966, IEEE Transactions on Antennas and Propagation, 14, 302

\bibitem[Zrake \& East(2016)]{ZraEas16}
Zrake, J., \& East, W.~E.\ 2016, ApJ, 817, 89

\bibitem[Zrake(2015)]{Zra15}
Zrake, J.\ 2015, arXiv:1512.05426

\end{thebibliography}
\end{document}